\documentclass[aps,prl,reprint,groupedaddress]{revtex4-2}

\usepackage{amsmath}
\usepackage{amsfonts}
\usepackage{graphicx}
\usepackage{xcolor}
 \usepackage{amssymb}

\begin{document}

\title{Programmable all-to-all entangling gates for cavity-coupled atoms}

\author{Angus Crookes}
\author{Ben Yuen}
\author{Angela Demetriadou*\thanks{a.demetriadou@bham.ac.uk}}

\affiliation{School of Physics and Astronomy, University of Birmingham, Edgbaston, Birmingham, B15 2TT, United Kingdom}

\date{\today}

\begin{abstract}
Fault tolerant quantum computers greatly benefit from having both high fidelity entangling gates and a non-local qubit connectivity. However, most entangling gates for neutral atoms are local, requiring separated atoms to be moved using optical tweezers to yield entanglement between them.  Here, we design a high fidelity M\o lmer S\o rensen-like entangling gate between neutral atoms coupled to a photonic cavity, enabling a programmable all-to-all connectivity through the shared cavity coupling. The gate is both robust to changes in cavity photon number, and enables selective entanglement between any subset of separated atoms through local laser addressing. This eliminates the need to physically move atoms to perform the gate, and therefore motional errors and erasure are suppressed. Using experimentally realistic parameters we demonstrate gate fidelities exceeding $F > 0.99$, with higher fidelities achievable through optimal control of the pump amplitudes.
\end{abstract}


\maketitle

\section{\label{introduction} Introduction}

Fault tolerant quantum computers must actively detect and correct errors faster than they occur, which requires entangling gates with low error rates and high fidelities \cite{nielsen2010quantum, lidar2013quantum, terhal2015quantum}. The ability to entangle arbitrary pairs of qubits also enables the implementation of quantum error correcting codes beyond those used in systems with only nearest neighbour entangling gates, such as the surface code \cite{stephens2014fault,andersen2020repeated,krinner2022realizing,google2023suppressing,google2025quantum}.  In particular,  quantum low-density-parity-check (qLDPC) codes often require non-local entangling gates, but are desirable because they have the potential to achieve error rates comparable to the surface code, but with a smaller encoding rate \cite{babar2015fifteen, breuckmann2021quantum, baspin2022connectivity, rakovszky2023physics}.

In this regard, ion trap qubits are one of the leading quantum computing platforms due to their excellent single and two qubit gate fidelities \cite{haffner2008quantum, ballance2016high, bruzewicz2019trapped}. For example,  single qubit gates with trapped ions have achieved error rates on the order of $10^{-6}$ \cite{leu2023fast} and $10^{-7}$ \cite{smith2025single}, while two qubit gates have error rates on the order of $10^{-3}$\cite{clark2021high} and $10^{-4}$ \cite{gaebler2016high}. The high fidelity entanglement is achieved with M\o lmer S\o rensen (MS) gates \cite{sorensen1999quantum, sorensen2000entanglement}, where ions, coupled through a shared vibrational mode, are driven with red and blue detuned laser fields. This interaction entangles the ions, but leaves the vibrational state separable at the end of the gate. It is therefore also robust to changes in the vibrational state of the ions, enabling high fidelity entanglement without requiring ground state cooling \cite{cirac1995quantum, schmidt2003realization, cai2023entangling}. Low error-rates and excellent experimental performance have made the MS gate the standard entangling operation in trapped-ion processors, and have also inspired variants in other qubit platforms \cite{mitra2020robust, mitra2023neutral, takahashi2017molmer, wang2021resilient}.


However, while MS gates generate high-fidelity entanglement, this process requires the ions to share a common collective vibrational mode. Implementing entangling gates between distant, or arbitrary pairs of ions in large-scale processors therefore requires shuttling ions between interaction zones where the gates are performed \cite{kielpinski2002architecture, pino2021demonstration, ransford202698}. This increases  hardware complexity, circuit depth, and introduces additional sources of decoherence and operational errors \cite{ballance2017high, kaushal2020shuttling, schoenberger2024shuttling, schier2026reinforcement}. Similar challenges also arise in other qubit platforms, such as Rydberg atom arrays, where entangling gates are also performed locally \cite{adams2020rydberg, wu2021concise}. In this case, atoms are dynamically rearranged using optical tweezers to bring selected qubits in, and out of, the interaction zone \cite{evered2023high, computing2026quantum}. However, repeated atom transport can introduce motional excitation and increase the probability of atom loss, resulting in erasure errors. In addition, quantum algorithms often require many gates and error correction steps, resulting in slow computation times due to the long transport process, which can in turn create further errors \cite{schoenberger2025shuttling}.

In this work, we derive a high fidelity M\o lmer S\o rensen-like gate for three-level $\Lambda$ atoms coupled to a photonic cavity, enabling entanglement between spatially separated qubits without physical transport or shuttling. In particular, any subset of atoms can be entangled through local laser addressing, with all other atoms acting as spectators to the gate. Our scheme therefore combines the robustness of M\o lmer S\o rensen gates with a simple neutral atom structure, enabling selective entanglement and high atom connectivity without encountering moving or shuttling errors.

\section{M\o lmer S\o rensen Hamiltonian}

Here, we consider an ensemble of $N$ atoms coupled to a photonic cavity, with a subset $n$ of the atoms driven by two external laser fields as shown in Figure \ref{fig:system_schematic} (a). All atoms are identical, with three levels in a $\Lambda$ configuration as shown in Figure \ref{fig:system_schematic} (b). The two lower levels $|g\rangle$ and $|e\rangle$ form a qubit, while the upper level $|r\rangle$ mediates coupling between the two lower states via cavity assisted Raman transitions. The coupling strengths for the $|g\rangle \leftrightarrow |r\rangle$  and for the $|e\rangle \leftrightarrow |r\rangle$ transitions are given by ($g_0$ and $\Omega_0$) and ($g_1$ and $\Omega_1$) respectively where $g$ denotes the cavity coupling and $\Omega$ the laser coupling. The full Hamiltonian describing this system is given by:

\begin{align}
\begin{split}
	H &= \omega_c a^{\dag}a + \sum_{i=1}^N\omega_{\text{eg}}|e\rangle^{(i)}\langle e|^{(i)} + \omega_{\text{rg}}|r\rangle^{(i)}\langle r|^{(i)}  \\ 
	& +\sum_{i=1}^N\left[g_0|r\rangle^{(i)}\langle 0|^{(i)}a + g_1|r\rangle^{(i)}\langle1|^{(i)}a + \text{h.c.}\right] \\ 
	&+ \sum_{i \in n} \left[\tilde{\Omega}_0^{(i)}|r\rangle^{(i)}\langle 0|^{(i)} + \tilde{\Omega}_1^{(i)}|r\rangle^{(i)}\langle 1|^{(i)} + \text{h.c.} \right]
	\label{full_ham}
\end{split}
\end{align}
\vspace{3mm}

\noindent where $\omega_{\text{eg}}$ and $\omega_{\text{rg}}$ define the atoms energy levels, $\omega_c$ is the frequency of the cavity mode, and $a^{\dag}$ and $a$ are the usual creation and annihilation operators of the field. The time harmonic pump amplitudes are given by $\tilde{\Omega}_0^{(i)} = \Omega_0^{(i)}e^{-i\omega_{L0}^{(i)}t}$ and $\tilde{\Omega}_1^{(i)} =\Omega_1^{(i)}e^{-i\omega_{L1}^{(i)}t}$ respectively, with time independent amplitudes $\Omega_0$ and $\Omega_1$. Importantly, we choose to only pump a subset of the atoms, which may be close or spatially separated from each other as shown in Figure \ref{fig:system_schematic} (a). 

To derive an entangling gate for the atoms, we first perform a series of unitary rotations, then adiabatically eliminate the upper level $|r\rangle$, define the laser frequencies $\omega_{L0} = \omega_c + \omega_{\text{eg}} + \delta$ and $\omega_{L1} = \omega_c - \omega_{\text{eg}} + \delta$ where $\delta$ is a small detuning parameter, and finally perform a rotating wave approximation to remove fast oscillations (see Appendix A for full derivation). In doing so, we derive a Hamiltonian of the form:



\begin{align}
	H_{\text{MS}} &=  g_{\text{eff}}\left(a^{\dag}e^{-i\delta't} + ae^{i\delta't}\right)S_x
	\label{molmer_sorensen_ham}
\end{align}
\vspace{3mm}

\noindent where $g_{\text{eff}} = - 2g\Omega/\Delta$ is the effective coupling strength,  $\delta' = \delta + Ng^2/\Delta$ is a detuning parameter, and $S_x = \sum_{i\in n} \sigma^{(i)}_x / 2$ is a sum only over the atoms that are driven by the external laser fields. Therefore, eq. (\ref{molmer_sorensen_ham}) is a gate operating only on the pumped atoms. Initially, for simplicity in deriving eq. (\ref{molmer_sorensen_ham}) we have taken $g_0 = g_1 = g$ and $\Omega_0 = \Omega_1 = \Omega$, and we evaluate this approximation on the gate fidelity later.  Importantly, the Hamiltonian in eq. (\ref{molmer_sorensen_ham}) is a MS gate. It has unitary evolution operator given by:

\begin{figure}[htbp]
	\centering
	\includegraphics[width=0.45\textwidth]{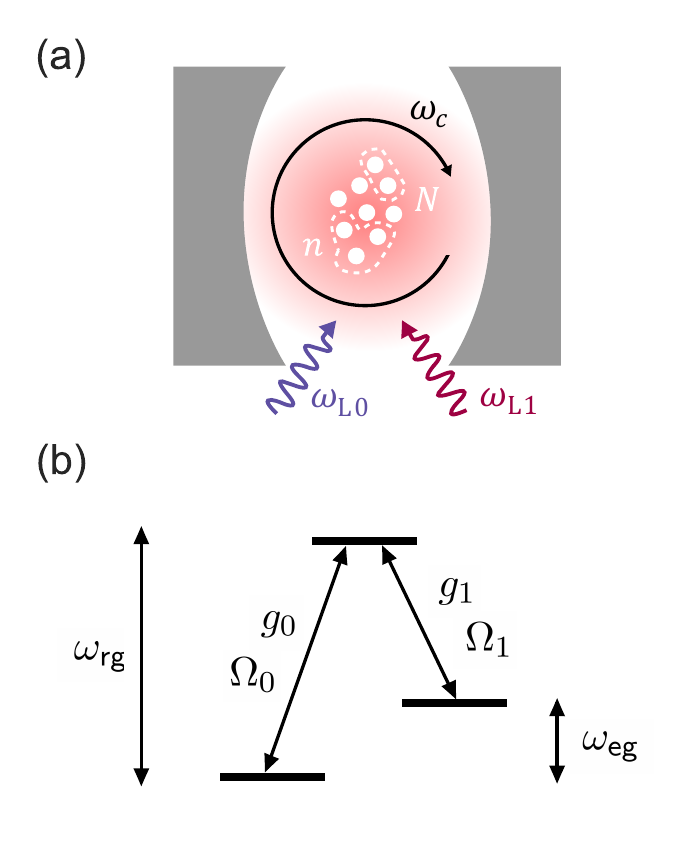}
	\caption{\textbf{Photonic cavity and atomic energy level diagrams.} \textbf{a} Schematic showing $N$  atoms coupled to a single mode of a photonic cavity; a subset $n$ of the atoms are also driven by a bi-chromatic laser (white dashed line). The atoms are not required to be local to one another, but must be simultaneously illuminated. \textbf{b} Energy level diagram of the atoms. Each atom is identical and has three levels in a $\Lambda$ configuration, with energy separation $\omega_{\text{rg}}$ between the upper excited state $|r\rangle$ and the lower ground state $|g\rangle$ and $\omega_{\text{eg}}$ between the lower excited and ground states $|e\rangle$ and $|g\rangle$ respectively. The transitions between  $|r\rangle \leftrightarrow |g\rangle$ from the cavity and pump occur with coupling rates $g_0$ and $\Omega_0$ respectively, and for $|r\rangle \leftrightarrow |e\rangle$ occur with rates $g_1$ and $\Omega_1$ respectively.  
	}
	\label{fig:system_schematic}
\end{figure}

\begin{equation}
	U_{\text{MS}} = e^{-i\left((\alpha(t)a^{\dag} + \alpha^*(t)a)S_x\right)}e^{i\beta(t)S_x^2}
	\label{molmer_sorensen_unitary}
\end{equation}
\vspace{3mm}

\noindent where $\alpha(t) = -i g_{\text{eff}} \left(1-e^{-i\delta't}\right)/\delta'$ and $\beta(t) = g_{\text{eff}}^2 \left(\delta' t - \sin \delta't\right)/\delta'^2$. The importance, and usefulness, of $U_{\text{MS}}$ comes from its insensitivity to the photon number state of the cavity. In particular, at time $T = 2\pi/\delta'$ the parameter $\alpha(T) = \alpha^*(T) = 0$. Therefore, $U_{\text{MS}}(T)$ becomes independent of $a$ and $a^{\dag}$, and at time $T$ the cavity mode is unchanged from its initial state, with the unitary evolution explicitly given by $U_{\text{MS}} = \exp(i\beta(T)S_x^2)$. This is an atom-atom interaction,  and only changes the atomic degrees of freedom. In addition, by choosing the detuning such that $|\delta'| = 2|g_{\text{eff}}|$ the phase becomes $\beta(T) = - \text{sgn}(\delta') \pi / 2$, and $U_{\text{MS}}$ becomes an operator used to generate maximally entangled states. 

\begin{figure*}[htbp]
	\centering
	\includegraphics[width=0.9\textwidth]{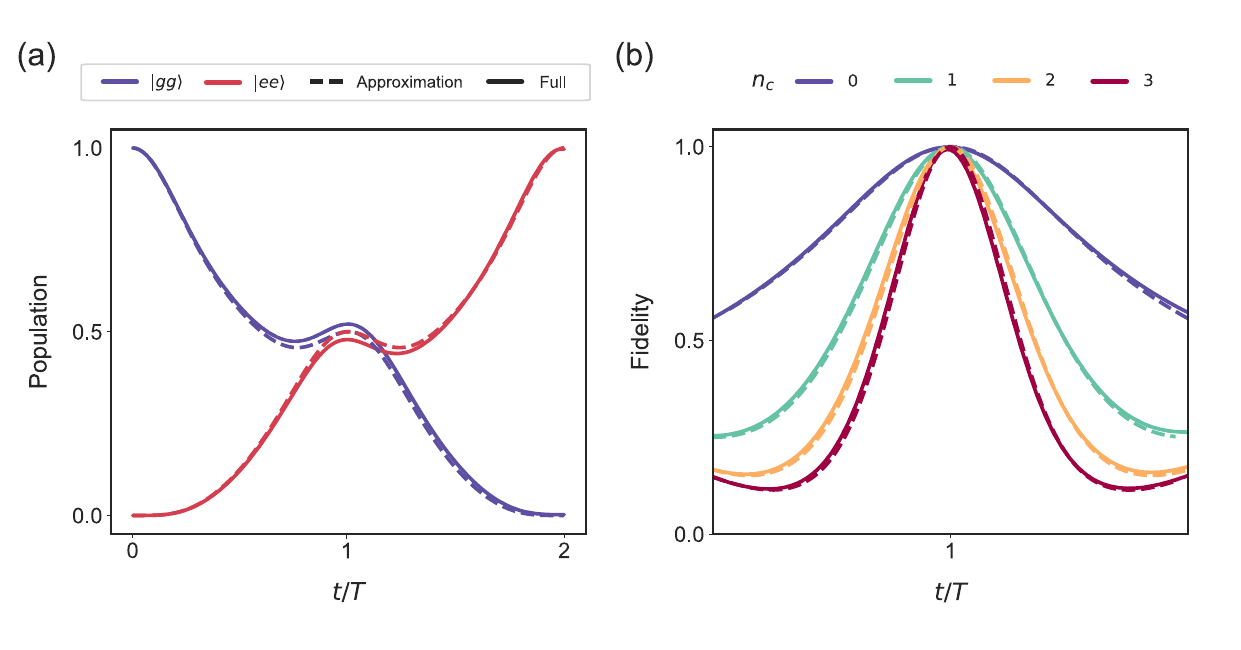}
	\caption{\textbf{M\o lmer S\o rensen-like entangling gate.} \textbf{a} Populations of the $|gg\rangle$ and $|ee\rangle$ states for a system containing two atoms ($N=2$) both driven by an external laser ($n=2$). The populations are calculated using the full Hamiltonian $H$ (solid lines) and with the approximate Hamiltonian $H_{\text{MS}}$ (dashed lines). The parameters used are $\omega_c = 1$, $\omega_{\text{rg}} = 1.6$, $\omega_{\text{eg}} = 0.01$, $g_0 = g_1 = 0.005$, and $\Omega_0 = \Omega_1 = 0.005$. $T$ is the time at which the entangling gate is complete using the M\o lmer S\o rensen Hamiltonian $T = 2\pi / \delta'$. \textbf{b} Entangled state fidelities $F=|\langle \psi_0|\psi(t)\rangle|^2$ for different initial photon states $|n_c\rangle|gg\rangle$. The fidelity is robust to changes in the photon number. }
	\label{fig:ms_gate}
\end{figure*}

For our system, the condition $|\delta'| = 2|g_{\text{eff}}|$ constrains the laser detuning parameter $\delta$ to be a solution to the equation $\delta = -(Ng^2 + 4g\Omega)/(\omega_{\text{rg}}-\omega_{c}-\omega_{\text{eg}}-\delta)$, which in turn defines both the pump frequencies $\omega_{L0}$ and $\omega_{L1}$. For two atoms ($N=2$) both driven by the external laser the $U_{\text{MS}}$ gate performs the following transformations:


\begin{align}
	\begin{split}
	|\phi_1\rangle &= |gg\rangle \rightarrow \frac{1}{\sqrt{2}}\left(|gg\rangle \mp i|ee\rangle\right) \\
	|\phi_2\rangle &= |ge\rangle \rightarrow \frac{1}{\sqrt{2}}\left(|ge\rangle \mp i|eg\rangle\right) \\
	|\phi_3\rangle &= |eg\rangle \rightarrow \frac{1}{\sqrt{2}}\left(\mp i|ge\rangle \mp i|eg\rangle\right) \\
	|\phi_4\rangle &= |ee\rangle \rightarrow \frac{1}{\sqrt{2}}\left(\mp i|gg\rangle \mp i|ee\rangle\right) 	
	\end{split}	
\end{align}
\vspace{3mm}

\noindent where $\mp$ denotes the sign of $-\text{sgn}(\delta')$. This transformation is equivalent to the controlled NOT gate up to a single qubit unitary rotation \cite{takahashi2017molmer}.

To demonstrate this, we first consider a system consisting of two atoms ($N=2$) both initially in the zero photon ground state $|0\rangle|gg\rangle$ and both driven by an external laser $(n=2)$. In Figure \ref{fig:ms_gate} (a) the populations of the $|gg\rangle$ and $|ee\rangle$ states are calculated using the full Hamiltonian $H$ (solid lines) and the approximate Hamiltonian $H_{\text{MS}}$ (dashed lines) using the parameters $\omega_c = 1$, $\omega_{\text{rg}} = 1.6$, $\omega_{\text{eg}} = 0.01$, $g_0 = g_1 = 0.005$, and $\Omega_0 = \Omega_1 = 0.005$ respectively. In both cases, the system is driven into the maximally entangled two qubit state at time $T$ given by $|\psi(T)\rangle = |\phi_1\rangle = |0\rangle\left(|gg\rangle + i|ee\rangle\right)/\sqrt{2}$. The fidelity of the entangled state in the simulation is calculated to be $F=0.999$. This is significantly above the $\approx 99\%$ threshold required for fault tolerant quantum computation, and demonstrates the validity of our approximation. Importantly, we also note that while the frequencies and coupling strengths were selected to give a state with high fidelity and short simulation time, high fidelity entangled states are possible over a wide range of parameters as discussed in Appendix B.

A significant feature of the MS gate, is its insensitivity to the bosonic state of the cavity. This is shown in Figure \ref{fig:ms_gate} (b) where we initialise the system in the state $|n_{\text{c}}\rangle|gg\rangle$,  where $n_c$ is the initial photon number, and determine the resulting entangled state fidelity. In all cases, the entangled state produced at time $T$ is $|\psi(T)\rangle = |n_{\text{c}}\rangle\left(|gg\rangle + i|ee\rangle\right)/\sqrt{2}$ with high fidelity. In particular, the maximum fidelity achieved for $n_{\text{c}}=3$ is $F = 0.993$ demonstrating the gates robustness against changes in the cavities photon number occupation. Therefore, in a similar fashion to the MS gates for trapped ion qubits, here the MS gate protects against errors due photon number fluctuations in experiments, and computations involving multiple gates. 

\begin{figure*}[htbp]
	\centering
	\includegraphics[width=0.9\textwidth]{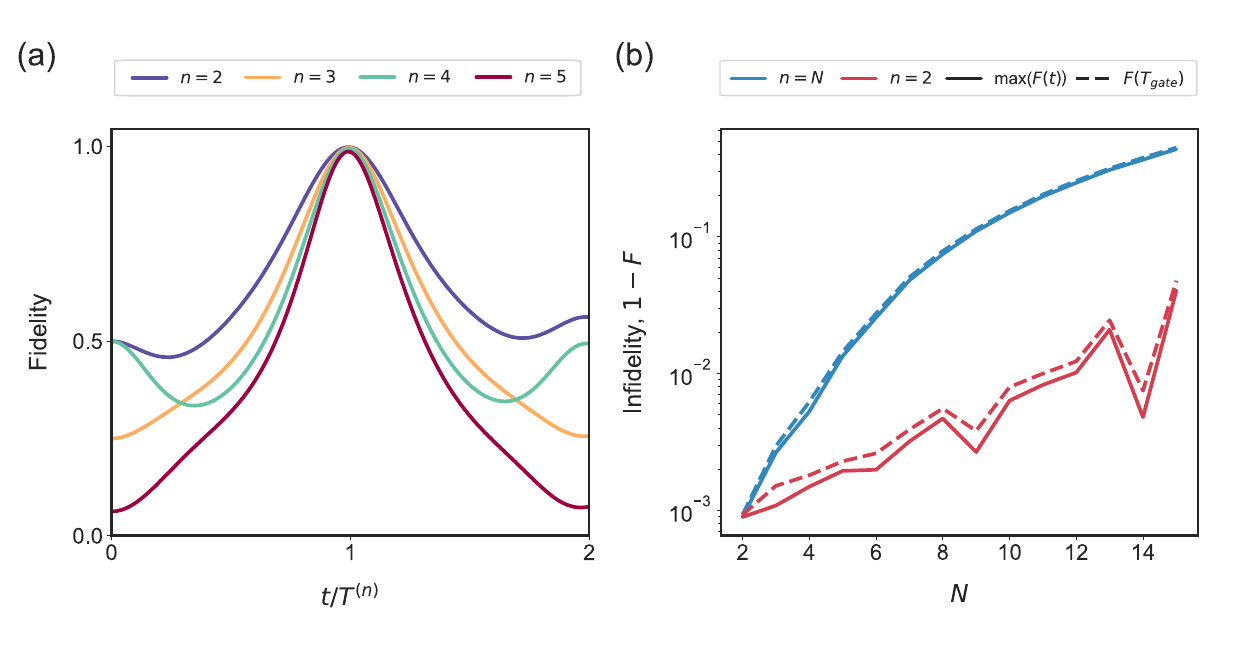}
	\caption{\textbf{M\o lmer S\o rensen-like entangling gate on a subset of atoms.} \textbf{a} Entangled state fidelities $F=|\langle \psi_0|\psi(t)\rangle|^2$ for a system containing five atoms ($N=5$) where only a subset $n$ of the atoms are driven by an external laser. The parameters used are $\omega_c = 1$, $\omega_{\text{rg}} = 1.6$, $\omega_{\text{eg}} = 0.01$, $g_0 = g_1 = 0.005$, and $\Omega_0 = \Omega_1 = 0.005$, and $T = 2\pi / \delta'$ is entangling gate time. In all cases, the subset of atoms excited by the laser are driven into an entangled state with a fidelity $F \geq 0.99$ \textbf{b} Infidelity ($1-F$) as a function of the total number of atom with the number of pumped atoms equal to $n=N$ (blue) and $n=2$ (red). The fidelity is calculated at the maximum (solid lines) and at the expected gate time (dashed lines).}
	\label{fig:selective_entanglement}
\end{figure*}

Importantly, our MS gate generates entangled states between any subset $n$ of the total atoms $N$ interacting with the cavity. Notably, the only atoms that become entangled are those for which the pump amplitudes $\Omega_0^{(i)}, ~\Omega_1^{(i)}$ are non-zero (i.e. the subset of $n$ atoms). This can be seen by the sum in eq. (\ref{molmer_sorensen_ham}) where $S_x$ contains a sum over $i\in n$ i.e. the subset of pumped atoms. Despite this, the condition for the entangling gate $|\delta'| = 2|g_{\text{eff}}|$ gives a detuning parameter $\delta = -(N+4)g^2/\Delta$ which depends on the total number of atoms interacting with the cavity.

In particular, for a subset $n \subset N$ atoms initially in the state  $|g\rangle^{\otimes N}$, the entangled states generated 
depend on whether $n$ is even or odd. In the subspace of the pumped atoms, the entangled states take the form:

\begin{equation}
	|\psi\rangle =
	\begin{cases}
		\frac{1}{\sqrt{2}}\left(|g\rangle^{\otimes n} + i^{n+1}\text{sgn}(\delta')|e\rangle^{\otimes n}\right) & n_{\text{even}} \\
		\frac{1}{\sqrt{2}}\left[\left(\frac{|g\rangle+i|e\rangle}{\sqrt{2}}\right)^{\otimes n}+\left(\frac{|g\rangle-i|e\rangle}{\sqrt{2}}\right)^{\otimes n}\right] & n_{\text{odd}}
	\end{cases}
	\label{multi_atom_entangled_states}
\end{equation} 
\vspace{3mm}

\noindent where for even numbers of atoms $n=n_{\text{even}}$, we create a maximally entangled GHZ state, and for odd numbers of atoms, $n=n_{\text{odd}}$, we create an entangled superposition state between $|g\rangle^{\otimes n}$ and states with an even number of excited atoms (see Appendix C for the full derivation). For example, for $n=3$ the entangled state generated is  $|\psi_T\rangle = \frac{1}{2}(|ggg\rangle - |gee\rangle - |ege\rangle - |gee\rangle)$ while for $n=4$ we get the maximally entangled state $|\psi_T\rangle = \frac{1}{\sqrt{2}}(|gggg\rangle + i\text{sgn}(\delta')|eeee\rangle$. 

Figure \ref{fig:selective_entanglement} (a) shows entanglement generation between a subset $n$ of $N=5$ atoms. The fidelity is calculated as $F = \langle \psi_{n} |\psi(T)\rangle|^2$ where $|\psi_n\rangle$ is given in eq.(\ref{multi_atom_entangled_states}). For all $n$ we generate an entangled state with a high fidelity. In particular, for $n=4$ and $n=5$ the maximum fidelities are $F = 0.995$ and $F = 0.986$ respectively. Therefore, only a small reduction in fidelity occurs when creating larger multi-atom entangled states. This is further demonstrated in \ref{fig:selective_entanglement} (b) where the infidelity is shown as a function of the total number of atoms $N$, with the number of pumped atoms given by $n=N$ (blue lines) and $n=2$ (red lines). The fidelity is also calculated at both the maximum (solid lines) and at the expected gate time (dashed lines). In both cases, the errors are exacerbated as the system size increases, due to an increase to small inaccuracies in the adiabatic and rotating wave approximations. However, even with a large number of atoms in the system, the two atom entangling gate still achieves a high fidelitiy with  $F>0.9$. Further details on the calculations shown in Figure \ref{fig:selective_entanglement} (b) can be found in Appendix D.

\begin{figure*}[htbp]
	\centering
	\includegraphics[width=1\textwidth]{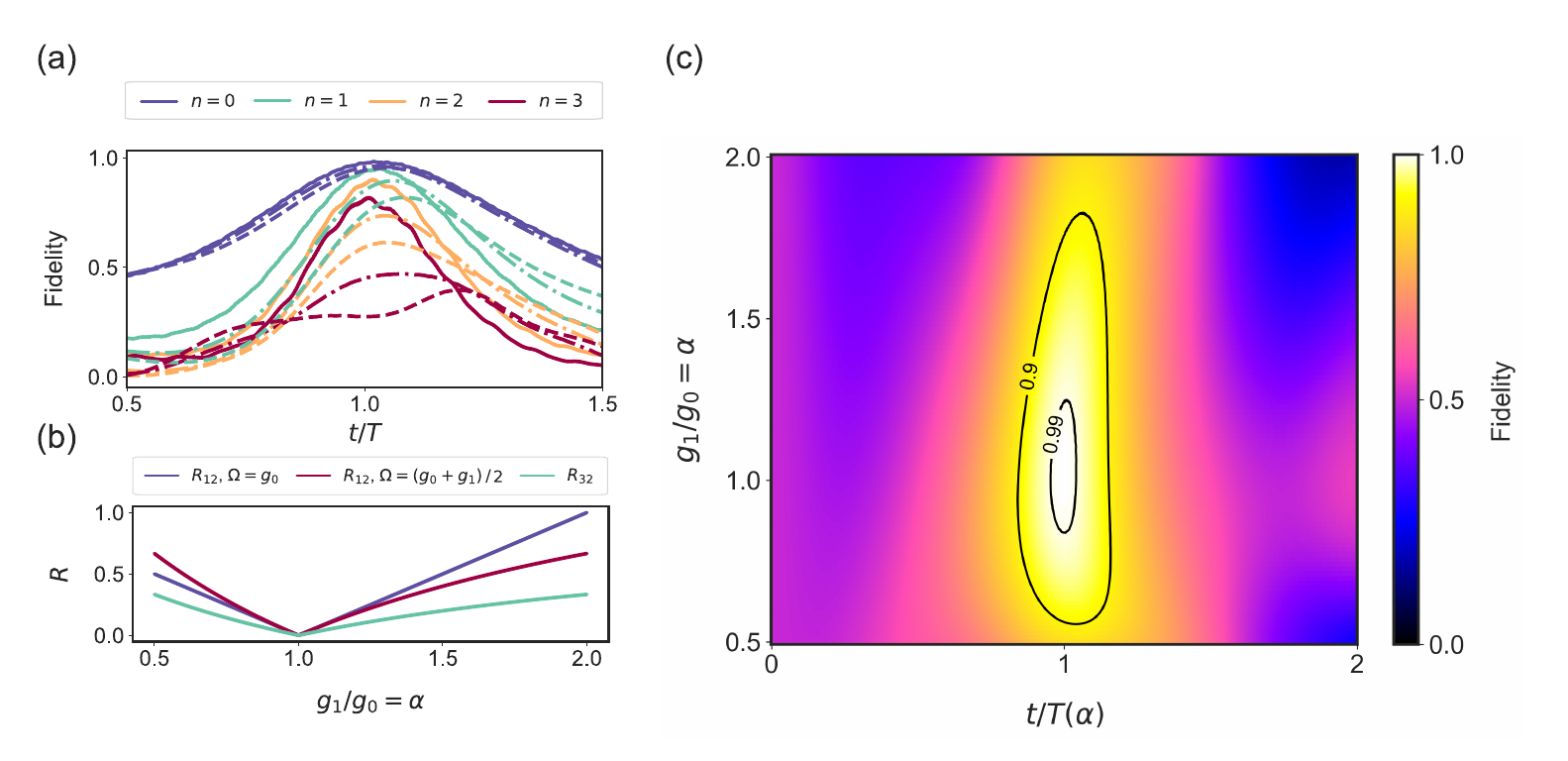}
	\caption{\textbf{Entanglement generation under unequal coupling to the photonic mode.} \textbf{a} Fidelity $F$ for a system containing two identical atoms ($N=2$) which both have coupling strengths defined by $g_1 = \alpha g_0$ where $\alpha = 1.5$. The simulation parameters used are $\omega_c = 1$, $\omega_{\text{rg}} = 1.6$, $\omega_{\text{eg}} = 0.01$, and $g_0 = 0.005$, and $\Omega_0 = \Omega_1 = \Omega$, and $T(\alpha) = 2\pi / \delta'(\alpha)$ is entangling gate time. The fidelity is calculated for a number of pump amplitudes including: $\Omega = g_0$ (dashed line) $\Omega = (g_0+g_1)/2$ (dash-dot line) and the optimal amplitude $\Omega_{\text{opt}}$ (solid line) for different initial photon states $|n_i\rangle|gg\rangle$. \textbf{b} Significance of each term in the Hamiltonian defined by the ratio $R_{ij} = H_i/H_j$ for different pump amplitudes $\Omega$. $R_{32}$ does not depend on the pump amplitude. \textbf{c} Fidelity $F$ as a function of time $t/T(\alpha)$ and coupling ratio $\alpha$ for a fixed pump amplitude $\Omega = (g_0 + g_1)/2$ starting in the initial state $|0\rangle|gg\rangle$. Contour lines for the regions of high fidelity $F=0.99$ and $F=0.9$ are also shown.    }
	\label{fig:unequal_coupling}
\end{figure*}

Unlike most trapped-ion and neutral-atom quantum computing architectures, which rely on local entangling gates and ion shuttling or atom rearrangement to implement interactions between distant qubits, our scheme exploits a shared photonic cavity mode that mediates long range entangling operations. Consequently, arbitrary subsets of atoms can be selectively entangled through local laser addressing without requiring physical transport of the qubits. Eliminating transport operations significantly simplifies hardware requirements to allow for miniaturisation, reduces circuit execution time, and avoids the accumulation of errors associated with repeated shuttling protocols. This capability also provides a promising route towards highly connected quantum processors capable of efficiently implementing advanced quantum error correcting codes, such as qLDPC codes.

\section{Sources of Errors}

When $g_0 \neq g_1$ the MS gate deviates from the ideal entangling gate operation. In particular, the general Hamiltonian for unequal coupling strengths (see Appendix A) is: 

\begin{align}
	\begin{split}
		H_{\text{MS}} &= -\frac{1}{\Delta}\sum_{i=1}^N \left(\frac{g_1^{2}-g_0^{2}}{2}\right)a^{\dag}a\sigma_z^{(i)} \\
		&+ g_{\text{eff}}^{(x)} \left(a^{\dag}e^{-i\delta't} + ae^{i\delta't}\right)S_x \\
		&+ g_{\text{eff}}^{(y)} \left(a^{\dag}e^{-i\delta't} - ae^{i\delta't}\right)S_y 
	\end{split}
	\label{full_molmer_sorensen_ham_main}
\end{align}
\vspace{3mm}


\noindent where $g_{\text{eff}}^{(x)} = -(g_0 + g_1)\Omega / \Delta$ and $g^{(y)}_{\text{eff}} = i(g_1-g_0)\Omega/\Delta$ describe effective coupling strengths for orthogonal rotations. The first term represents an AC-Stark shift, which for unequal coupling strengths produces a state dependent frequency shift that changes the internal state of the atom depending on the occupation number of the cavity. The third term represents an additional orthogonal atom dependent force along the $S_y$ axis. Both terms, which we denote $H_{\text{MS}}^{(1)}$ and $H_{\text{MS}}^{(3)}$ respectively, degrade the fidelity of the entangled state.

To determine the effect these additional terms have on the entanglement, in Figure \ref{fig:unequal_coupling} we again consider two atoms coupled to a cavity mode and both pumped by external lasers. The atoms are initialised in the zero photon ground state $|0\rangle|gg\rangle$ and to model a general coupling strength we assume the $|e\rangle, |g\rangle \leftrightarrow  |r\rangle$  coupling strengths are defined by the ratio $g_1/g_0 = \alpha$. In Figure \ref{fig:unequal_coupling} (a) we plot the fidelity of the entangled state for a large difference in coupling strengths, $\alpha = 1.5$, and for three different choices of pump amplitude $\Omega_0=\Omega_1=\Omega$: (1) $\Omega = g_0$ (dashed lines) and (2) $\Omega = (g_1 + g_0) / 2$ (dash-dotted lines) and (3) the optimal $\Omega_{\text{opt}}$ (solid lines) determined from a parameter sweep over a suitable range values. Note that in this case, the M\o lmer S\o rensen gate condition $\delta' = 2|g_{\text{eff}}|$ gives us the pump detuning $\delta$ through solving the equation $\delta = -(N(g_0^2 + g_1^2)/2 + 2\Omega(g_0+g_1))/(\omega_{\text{rg}}-\omega_{c}-\omega_{\text{eg}}-\delta)$, which in turn defines both the pump frequencies $\omega_{L0}$ and $\omega_{L1}$

In all three cases, the asymmetry in the cavity coupling strengths reduces the entangled states fidelity. The decrease in fidelity is larger for states with greater numbers of photons since the AC-Stark shift term is $\propto a^{\dag}a$. However, the fidelity can be controlled by optimising the pump amplitude. In particular, the ratio between coefficients of the AC Stark ($H_{\text{MS}}^{(1)}$) and MS ($H_{\text{MS}}^{(2)}$) terms in the Hamiltonian is given by $R_{12} = |(g_1^2 - g_0^2)/\Delta g_{\text{eff}}^{(x)}|$. In general, this is equal to $R_{12} = |\alpha-1|g_0 / \Omega$ and we can therefore tune the pump amplitude $\Omega$ for a given $\alpha$ to minimise the contribution of the Stark shift. The ration $R_{12}$ for the original pump amplitude $\Omega = g_0$ is shown in Figure \ref{fig:unequal_coupling} (b). However, since $R_{12}\propto \Omega^{-1}$ by defining $\Omega = (g_0 + g_1) / 2$ we get $R_{12} = 2|\alpha-1|/(\alpha+1)$ which suppresses the ratio whenever $\Omega > g_0$. This is also shown in Figure \ref{fig:unequal_coupling} (b) where for $\alpha > 1$ (i.e. $g_1 > g_0$) the contribution of the AC Stark shift term is reduced. In Figure \ref{fig:unequal_coupling} (a) the fidelity of the dash-dotted line represents this choice of pump amplitude, and is greater than the original pump amplitude.  

While it appears that increasing $\Omega$ increases the fidelity, eventually we break the rotating wave, and adiabatic approximations we make in the derivation of $H_{\text{MS}}$ and the fidelity will in-fact decrease. Therefore, for each $\alpha$ there is an optimal pump amplitude $\Omega_{\text{opt}}$ which maximises the fidelity, as shown in Figure \ref{fig:unequal_coupling} (a). This provides a high fidelity for all initial photon states $|n_c\rangle|gg\rangle$. For $n_c = 0$ the fidelity is $F = 0.98$.

Therefore, the choice of pump amplitude $\Omega$ greatly suppresses the AC stark shift, however, the ratio $R_{32} = |\alpha-1|/(\alpha+1)$ - which represents a ratio between the coefficients of the $S_y$ ($H_{\text{MS}}^{(3)}$) and $S_x$ ($H_{\text{MS}}^{(2)}$) gates in eq. (\ref{full_molmer_sorensen_ham_main}) - is independent of $\Omega$. Therefore, this term will always reduce the fidelity of the entangled states, irrespective of the pump amplitude. Nevertheless, $R_{32} < R_{12}$ (see Figure \ref{fig:unequal_coupling} (b)) and therefore has a significantly smaller impact than the Stark shift for our chosen pump amplitudes.

Finally, the fidelity as a function of the coupling ratio $\alpha$ is shown in Figure \ref{fig:unequal_coupling} (c) for a fixed pump amplitude $\Omega = (g_0+g_1)/2$. Notably, for moderate or small differences in the coupling strengths we still obtain high fidelities exceeding $F=0.99$. 

\begin{figure}[htbp]
	\centering
	\includegraphics[width=0.5\textwidth]{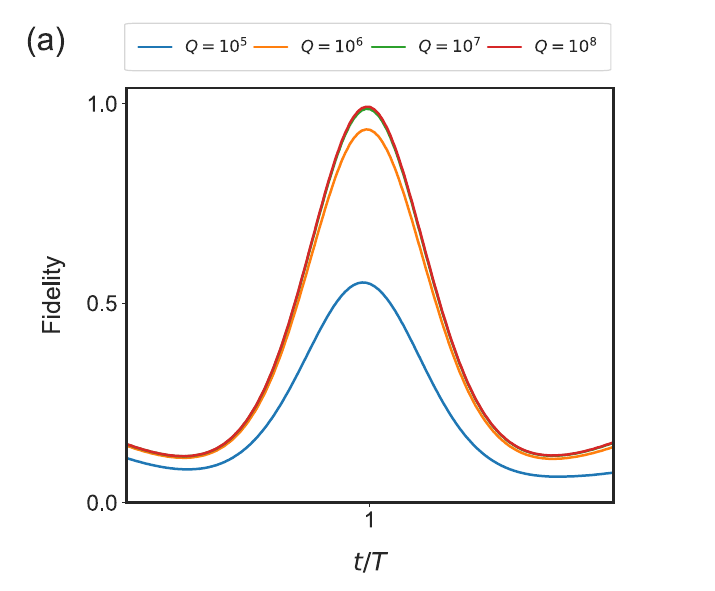}
	\caption{\textbf{Entangled state fidelity dependence on cavity finesse} \textbf{a} Entangled state fidelity for two atoms with cavity loss $\kappa = 1/2Q$ where $Q$ is the cavity quality factor.  The parameters used are $\omega_c = 1$, $\omega_{\text{rg}} = 1.6$, $\omega_{\text{eg}} = 0.01$, $g_0 = g_1 = 0.005$, and $\Omega_0 = \Omega_1 = 0.005$, and $T = 2\pi / \delta'$ is entangling gate time. }
	\label{fig:quality_factors}
\end{figure}


Another important source of error is cavity dissipation, which limits the coherence of the cavity-mediated interaction. This is quantified by the cavity quality factor $Q = \omega_c / 2\kappa$ where $\kappa$ is the cavity decay rate.To quantify the effect of cavity losses, we model the system using a standard Lindblad Master Equation:

\begin{equation}
	\dot{\rho} = -i\left[H, \rho\right] + \kappa \left(a\rho a^{\dag} - \frac{1}{2}\{ a^{\dag}a, \rho \} \right)
\end{equation}

\noindent where $\rho$ is the density matrix of the cavity-atom system. 

The resulting gate fidelities are shown in Figure \ref{fig:quality_factors} (a) for quality factors ranging from  $Q = 10^5$ to $10^8$. For the atom and pump parameters considered here, cavities with $Q = 10^7$ and $Q = 10^8$ achieve maximum fidelities of $F = 0.986$ and $F = 0.992$ respectively, demonstrating that high fidelity operation remains possible with experimentally accessible nanophotonic cavities. As expected, lower quality factors lead to a more pronounced reduction in fidelity, since photons are more likely to decay from the cavity during the gate operation time. 

Importantly, this reduction can be reduced. Since the effective interaction scales as $g_{\text{eff}} \propto g\Omega$ and the gate time scales as $T = \pi / g_{\text{eff}}$, increasing either the atom-cavity coupling strength or the pump amplitude reduces the interaction time, hence decreasing the probability photon loss during the gate. Furthermore, the pulse shapes through out this work have not been optimised. We therefore expect that applying quantum optimal control techniques to the driving fields will further suppress the effects of cavity dissipation and enable even higher gate fidelities.


\section*{Conclusion}

In conclusion, we have derived a high fidelity M\o lmer S\o rensen-like entangling gate for three level $\Lambda$ atoms coupled to a photonic cavity. Our scheme combines the robustness of the M\o lmer S\o rensen interaction with programmable long-range connectivity enabled by the cavity mode to entangle distant atoms. These results establish cavity mediated M\o lmer S\o rensen interactions as a promising route towards scalable, highly connected quantum processors, and motivate future experimental demonstrations of fault tolerant architectures based on long-range entangling gates.

\section*{Data Availability}


\section*{APPENDIX A: DERIVATION OF M\O LMER S\O RENSEN EFFECTIVE HAMILTONIAN} \label{appendix_a}

To derive the M\o lmer S\o rensen (MS) Hamiltonian in eq. (\ref{molmer_sorensen_ham}) we first perform a unitary rotation $U_1 = \exp\left(iH_1 t\right)$ of the full Hamiltonian in eq. (\ref{full_ham}) where

\begin{align}
	H_1 &= \omega_c a^{\dag}a + \sum_{i=1}^N\omega_{\text{eg}}|e^{(i)}\rangle\langle e^{(i)}| + \omega_{L0}|r^{(i)}\rangle\langle r^{(i)}|  
	\label{H2}
\end{align}
\vspace{3mm}

\noindent and the new Hamiltonian $H' = U_1HU_1^{\dag} - H_1$ is:

\begin{align}
	\begin{split}
	H' &= \Delta\sum_{i=1}^N |r\rangle^{(i)}\langle r|^{(i)} + \big[\sum_{i=1}^NM^{(i)}_0(t)|r\rangle^{(i)}\langle 0|^{(i)} \\
	&+ \sum_{i=1}^NM^{(i)}_1(t)|r\rangle^{(i)}\langle 1|^{(i)} + \text{h.c.}\big]
	\end{split}
\end{align}
\vspace{3mm}

\noindent where $\Delta = \omega_{\text{rg}} - \omega_{L0}$ and 

\begin{equation}
	M^{(i)}_0(t) = g^{(i)}_0e^{i(\omega_{L0}-\omega_c)t}a + \Omega^{(i)}_0
\end{equation}

\begin{equation}
	M^{(i)}_1 = g^{(i)}_1e^{i(\omega_{L0}-\omega_{10}-\omega_c)t}a + \Omega^{(i)}_1e^{i(\omega_{L0} - \omega_{10}-\omega_{L1})t}
\end{equation}
\vspace{3mm}

\noindent are operators that act on the $i^{\text{th}}$ atom. Note, we have not yet assumed any symmetry restrictions on $g_0$ and $g_1$. Now, we adiabatically eliminate states containing the upper atomic level $|r\rangle$. In general, the equation for the new Hamiltonian is given by $H_{\text{eff}} \approx PH'P - PH'Q(QH'_0Q)^{-1}QH'P$ where $H'_0 = \Delta \sum_{i=1}^N |r^{(i)}\rangle\langle r^{(i)}|$ and  $P$ is a projection operator onto the desired subspace

\begin{equation}
	P = \bigotimes_{i=1}^N (\mathbb{I}-|r^{(i)}\rangle\langle r^{(i)}|)
\end{equation} 
\vspace{3mm}

\noindent and $Q = \mathbb{I} - P$. The resulting Hamiltonian under adiabatic elimination is:

\begin{align}
	\begin{split}
	H_{\text{eff}} &= -\frac{1}{\Delta} \sum_{i=1}^N  M_0^{(i)\dag}M^{(i)}_0|g^{(i)}\rangle\langle g^{(i)}| \\ 
		&+ M_1^{(i)\dag}M^{(i)}_1|e^{(i)}\rangle\langle e^{(i)}| \\ &+ M_1^{(i)\dag}M^{(i)}_0|e^{(i)}\rangle\langle g^{(i)}| \\
		&+ M_0^{(i)\dag}M_1^{(i)}|g^{(i)}\rangle\langle e^{(i)}|
	\end{split}
\end{align}
\vspace{3mm}

\noindent and only acts on the qubit subspace $|e\rangle,~|g\rangle$. Next, we define the laser frequencies to be $\omega_{L0} = \omega_c + \omega_{\text{eg}} + \delta$ and $\omega_{L1} = \omega_c - \omega_{\text{eg}} + \delta$ where $\delta$ is a small detuning parameter. By assuming that $\delta << \omega_{\text{eg}}, ~ \omega_c$ we can make a rotating wave approximation on $H_{\text{eff}}$ resulting in a new Hamiltonian:

\begin{align}
	\begin{split}
	H_{\text{eff}} &= -\frac{1}{\Delta}\sum_{i=1}^N\left(g_0^{(i)2} a^{\dag}a + \Omega_0^{(i)2}\right)|g^{(i)}\rangle\langle g^{(i)}| \\
	&+ \left(g_0^{(i)2} a^{\dag}a + \Omega_1^{(i)2}\right)|e^{(i)}\rangle\langle e^{(i)}| \\
	&+ g_1^{(i)}\Omega_0^{(i)}e^{-i\delta t}a^{\dag} |e^{(i)}\rangle\langle g^{(i)}| \\
	&+ g_0^{(i)}\Omega_1^{(i)}e^{i\delta t}a |e^{(i)}\rangle\langle g^{(i)}| \\
	&+ g_{1}^{(i)}\Omega_0^{(i)}e^{i\delta t} a |g^{(i)}\rangle\langle e^{(i)}| \\
	&+ g_0^{(i)}\Omega_1^{(i)}a^{\dag}e^{-i\delta t}|g^{(i)}\rangle\langle e^{(i)}|
	\end{split}
\end{align}
\vspace{3mm}

\noindent which has both Jaynes-Cummings and anti-Jaynes-Cummings terms e.g. $a|e^{(i)}\rangle\langle g^{(i)}|$ and $a|g^{(i)}\rangle \langle e^{(i)}|$ respectively. Finally, we make another unitary transformation following $U_2 = \exp\left(-iH_2t\right)$ where

\begin{align}
	\begin{split}
	H_2 &= \frac{1}{2\Delta}\sum_{i=1}^N  g_0^{(i)2} a^{\dag} a |g^{(i)}\rangle\langle g^{(i)}| + g_1^{(i)2} a^{\dag} a |e^{(i)}\rangle\langle e^{(i)}| \\
	&+ \frac{1}{\Delta}\sum_{i=1}^N \Omega_0^{(i)2}|g^{(i)}\rangle\langle + \Omega_1^{(i)2}|e^{(i)}\rangle\langle e^{(i)}| 
	\end{split}
\end{align}
\vspace{3mm}

\noindent and, by assuming $g_0^{(i)} = g_0$ and $g_1^{(i)} = g_1$, defining

\begin{equation}
	\delta' = \delta + \sum_{i=1}^N \frac{g_0^{(i)2}+g_1^{(i)2}}{2\Delta}
\end{equation}
\vspace{3mm}

\begin{equation}
	\delta' = \delta + N (g_0^{2}+g_1^{2})/2\Delta
\end{equation}

\noindent and choosing to pump only a selection of the atoms,

\begin{equation}
	\Omega_0^{(i)} = \Omega_1^{(i)} = 
	\begin{cases}
		\Omega & i \in n \\
		0 &  $\text{otherwise}$
	\end{cases}
\end{equation}
\vspace{3mm}

\begin{figure*}[htbp]
	\includegraphics[width=1\textwidth]{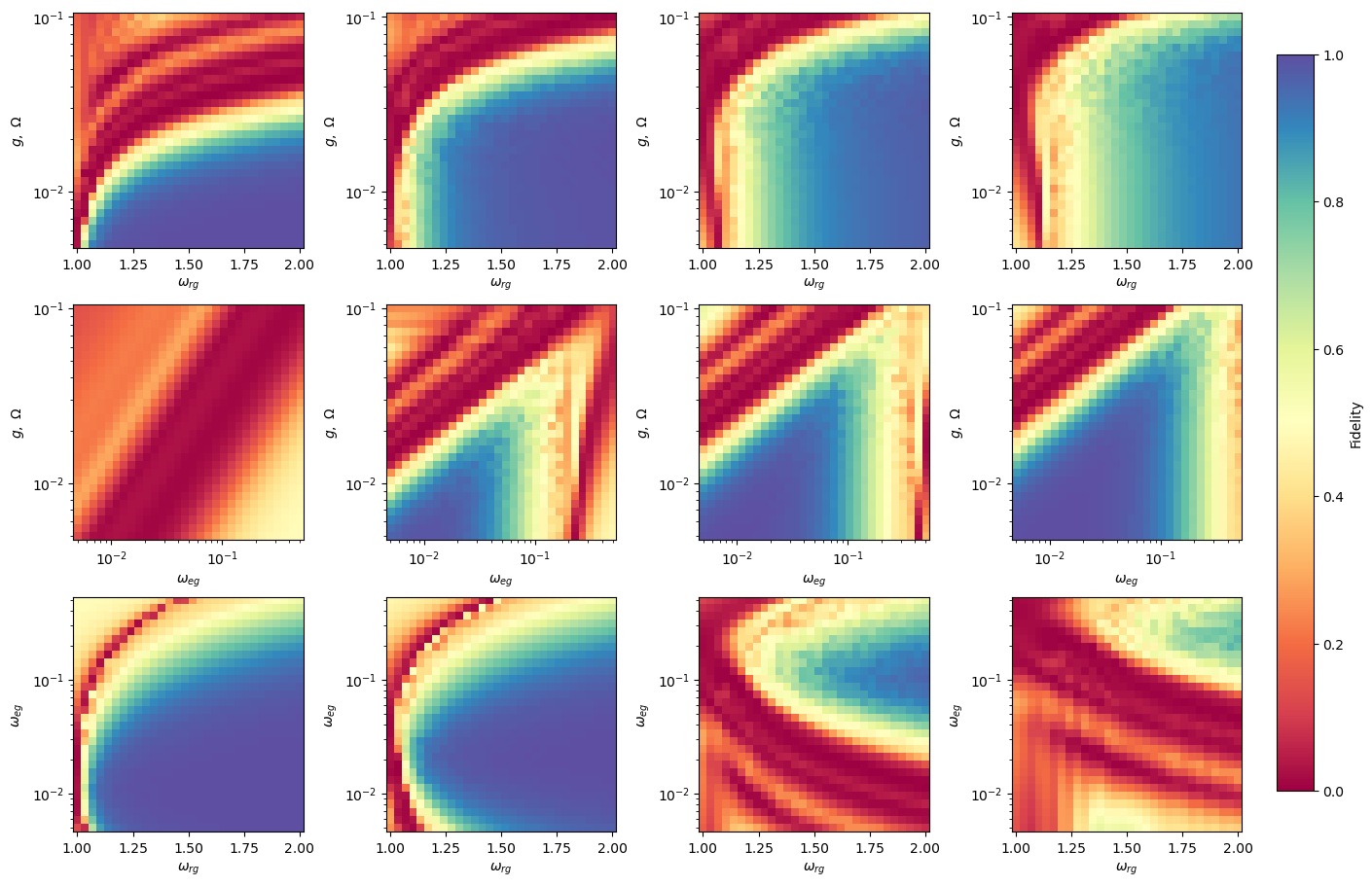}
	\caption{\textbf{Fidelity parameter sweep} Fidelity $F(T)$ evaluated at the expected gate time $T$ for a system containing two atoms ($N=2$) both driven by an external laser ($n=2$) in a zero loss cavity $\kappa = 0$. (TOP) Fidelity for fixed qubit detuning at $\omega_{eg} = 0.01, 0.05, 0.1, 0.15$ (left to right) as a function of upper level spacing $\omega_{rg}$ and coupling $g = \Omega $. (MIDDLE) Fidelity for fixed upper level spacing at $\omega_{rg} = 1, 1.2, 1.4, 1.6$ (left to right) as a function of lower level spacing $\omega_{eg}$ and coupling $g = \Omega $. (BOTTOM) Fidelity for fixed coupling  at $g = \Omega = 0.005, 0.01, 0.05, 0.1$ (left to right) as a function of lower and upper level spacings $\omega_{eg}$ and $\omega_{rg}$.}
	\label{fig:appendix_b}
\end{figure*}

\noindent we get the MS-like Hamiltonian:

\begin{align}
	\begin{split}
	H_{\text{MS}} &= -\frac{1}{\Delta}\sum_{i=1}^N \left(\frac{g_1^{2}-g_0^{2}}{2}\right)a^{\dag}a\sigma_z^{(i)} \\
	&+ g_{\text{eff}}^{(x)} \left(a^{\dag}e^{-i\delta't} + ae^{i\delta't}\right)S_x \\
	&+ g_{\text{eff}}^{(y)} \left(a^{\dag}e^{-i\delta't} - ae^{i\delta't}\right)S_y 
	\end{split}
	\label{full_molmer_sorensen_ham}
\end{align}
\vspace{3mm}

\noindent where $S_{x/y} = \sum_{i\in n}\sigma_{x/y}^{(i)} $ and $g_{\text{eff}}^{(x)} = - \Omega(g_0 + g_1) / \Delta$ and $g_{\text{eff}}^{(y)} = - i\Omega(g_0 - g_1) / \Delta$. Importantly eq. (\ref{full_molmer_sorensen_ham}) reduces to eq. (\ref{molmer_sorensen_ham}) under the equal coupling condition $g_0=g_1$. The first term represents an AC Stark shift term, which acts to reduce the fidelity of the entangled state when $g_0\neq g_1$. The third term is an orthogonal rotation to the MS rotation, which also reduces the fidelity. A discussion of the impact of these terms is included in the main text.

\section*{APPENDIX B} \label{appendix_b}

Here we demonstrate the validity of the approximations used in deriving eq. (\ref{molmer_sorensen_ham}) and how for a large range of parameters a high fidelity entangled state is generated. The parameter sweeps changing $\omega_{\text{rg}}$, $g$, $\Omega$ and $\omega_{\text{eg}}$ are shown in Figure \ref{fig:appendix_b}.

\section*{APPENDIX C} \label{appendix_c}

Here, we summarise the derivation of the states in eq. (\ref{multi_atom_entangled_states}) when $n$ atoms are pumped. The initial state for the $n$ pumped atoms can be written as $|g\rangle^{\otimes n} = \left(|+\rangle + |-\rangle\right)^{\otimes n}/\sqrt{n}$ where $|\pm\rangle$ are the usual eigenstates of the $\sigma_x$ operator. This can also be expressed as:

\begin{equation}
	|\psi(0)\rangle = \frac{1}{2^{n/2}}\sum_{\mathbf{s}}|\mathbf{s}\rangle
	\label{initial_state_pm_string}
\end{equation}
\vspace{3mm}

\noindent where  $\mathbf{s} = (s_1, s_2 ... s_n)$ and $s_j \in \{+, -\}$ is a chain of the $|\pm\rangle$ eigenstates. For a MS gate the unitary evolution operator at time $T=2\pi/\delta'$ becomes $U(T) = \exp\left(\frac{-\text{sgn}(\delta')\pi S_x^2}{2}\right)$. Applying this to eq. (\ref{initial_state_pm_string}) and using $S_x|\mathbf{s}\rangle = (n/2 - k)|\mathbf{s}\rangle$ where $k$ is the number of $-1$ eigenstates in the string $\mathbf{s}$ the time evolved state is:

\begin{equation}
	|\psi(T)\rangle = \frac{1}{2^{n/2}}\sum_{\mathbf{s}} \exp\left(\frac{-\text{sgn}(\delta')\pi}{2}\left(\frac{n}{2}-k\right)^2\right)|\mathbf{s}\rangle
	\label{time_evolved_general_state}
\end{equation}
\vspace{3mm}

\noindent \textbf{Even n:} To derive the entangled states for even $n$ we can take $n = 2r$ and defining $m = r - k$ we have:

\begin{equation}
	e^{-i\text{sgn}(\delta')\frac{\pi m^2}{2}} = \begin{cases}
		1 & $\text{if} $m$ \text{is even} $\\
		-i\text{sgn}(\delta') &  $\text{if} $m$ \text{is odd} $
	\end{cases}
\end{equation}
\vspace{3mm}

\noindent which is equivalent to: 

\begin{equation}
	e^{-i\text{sgn}(\delta')\frac{\pi m^2}{2}} = \frac{1-i}{2} + (-1)^{r+k}\frac{1+i}{2}
\end{equation}
\vspace{3mm}

\noindent This can then be substituted into eq. (\ref{time_evolved_general_state}) which up to a global phase gives:

\begin{equation}
	|\psi(T)\rangle = \frac{1}{\sqrt{2}}\left[|g\rangle^{\otimes n} + i^{n+1}\text{sgn}(\delta')|e\rangle^{\otimes n}\right]
\end{equation}
\vspace{3mm}

\noindent as stated in the main text.  
\newline

\noindent \textbf{Odd n:} To derive the entangled states for odd $n$ in eq. (\ref{time_evolved_general_state}) we instead take $n = 2r + 1$ where $r\in\mathbb{Z}$ and so $m = r - k + 1/2$ is a half integer. Therefore, one can write:

\begin{equation}
	e^{-i\pi\eta m^2/2} = \frac{e^{-i\pi\eta/8}}{\sqrt{2}}\left(e^{i m\pi/2} + e^{-i m\pi/2}\right)
\end{equation}
\vspace{3mm}

\noindent where $\eta = \text{sgn}(\delta')$.  Using this expression, one can then write the time evolved quantum state as

\begin{equation}
	|\psi(T)\rangle = \frac{1}{\sqrt{2}}\left[e^{i\pi S_x/2}|g\rangle^{\otimes n} + e^{-i\pi S_x/2}|g\rangle^{\otimes n} \right]
\end{equation}
\vspace{3mm}

\noindent where we have used $\sum_{\mathbf{s}}e^{i\pi m/2}|\mathbf{s}\rangle/2^{n/2} = e^{i\pi S_x/2} |g\rangle^{\otimes n}$ and ignored the global phase. Finally, writing the sum of the exponential $\sigma_x^{(j)}$ operators as a product, and using the identity $\exp(\pm i\pi\sigma^{(j)}_x/4) = (\mathbb{I} \pm i\sigma_x^{(j)})/\sqrt{2}$ gives:

\begin{equation}
	|\psi_T\rangle = \frac{1}{\sqrt{2}}\left[\left(\frac{|g\rangle+i|e\rangle}{\sqrt{2}}\right)^{\otimes n} + \left(\frac{|g\rangle-i|e\rangle}{\sqrt{2}}\right)^{\otimes n}\right]
\end{equation}
\vspace{3mm}

\noindent as stated in the main text.

\section*{APPENDIX D} \label{appendix_d}

It is computationally expensive to calculate the time evolution, and fidelity, of many particle entangled states. In particular, the dimension of eq. (\ref{full_ham}) is $n_{\text{cav}} \cdot 3^N$ where $n_{\text{cav}}$ is the cavity photon truncation number, and $N$ is the total number of atoms. Hence, for large atom numbers e.g. $N>6$  the exponential growth of the Hilbert space makes calculations expensive and slow. However, since the $n$ pumped atoms are indistinguishable, and the $N-n$ spectator atoms are indistinguishable, here we instead make a Jordan-Schwinger transformation \cite{jordan1935zusammenhang, schwinger1952angular}:

\begin{equation}
	\sum_{i=1}^N \sigma_{ab}^{(i)} \rightarrow c_a^{\dag}c_b + d_a^{\dag}d_b ~~~~~~ \sum_{i=1}^n \sigma_{ab}^{(i)} \rightarrow d_a^{\dag}d_b
\end{equation}
\vspace{3mm}

\noindent where the excitation of indistinguishable atoms, is described through bosonic creation and annihilation operators acting on number states. In particular $c_a^{\dag}$ describes the excitation of a spectator atom into the state $|a\rangle$ while $d_a^{\dag}$ describes excitation of a pumped atom into the state $|a\rangle$. In this picture, the total Hamiltonian in eq. (\ref{full_ham}) becomes: 

\begin{equation}
	\begin{aligned}
		H(t) =\;& \Delta\big(c_r^\dagger c_r + d_r^\dagger d_r\big)
		+ \Omega_0\big(c_r^\dagger c_0 + \text{h.c.}\big) \\[4pt]
		&+ g_0\Big(e^{\,i w_{10} t}\,a\,\big(c_r^\dagger c_0 + d_r^\dagger d_0\big) + \text{h.c.}\Big) \\[4pt]
		&+ g_1\Big(e^{\,i w_{11} t}\,a\,\big(c_r^\dagger c_1 + d_r^\dagger d_1\big) + \text{h.c.}\Big) \\[4pt]
		&+ \Omega_1\Big(e^{-i w_{12} t}\,c_r^\dagger c_1 + \text{h.c.}\Big)
	\end{aligned}
\end{equation}
\vspace{3mm}

\noindent where $\Delta = E_r - \omega_{L0}$, $w_{10} = \omega_{L0}-\omega_c$
$w_{11} = \omega_{L0}-E_e-\omega_c$, and $w_{12} = \omega_{L1}+E_e-\omega_{L0}$. In addition, in this picture, the total number of excitation within the atomic subspace is conserved. This truncates the total dimension of the Hamiltonian to: 

\begin{equation}
	n_{\text{cav}} \cdot \frac{(n-1)(n-2)}{2} \cdot \frac{(N-n-1)(N-n-2)}{2}
\end{equation}
\vspace{3mm}

\noindent which is much smaller than $n_{\text{cav}} \cdot 3^N$ and hence allows computation of atomic fidelities for atom numbers up to $N\approx 15$. 


%

\end{document}